\documentclass[sn-mathphys-num,oneside,referee]{sn-jnl}

\usepackage{graphicx}%
\usepackage{multirow}%
\usepackage{amsmath,amssymb,amsfonts}%
\usepackage{amsthm}%
\usepackage{mathrsfs}%
\usepackage[title]{appendix}%
\usepackage{xcolor}%
\usepackage{textcomp}%
\usepackage{manyfoot}%
\usepackage{booktabs}%
\usepackage{algorithm}%
\usepackage{algorithmicx}%
\usepackage{algpseudocode}%
\usepackage{listings}%
\usepackage{chemformula}
\usepackage{siunitx}
\usepackage[utf8]{inputenc}

\newcommand{\cvs}{\ensuremath{\mathrm{CsV_3Sb_5}} }
\newcommand{\tcdw}{$T_{\text{CDW}}$ }
\newcommand{\nscvs}{\ensuremath{\mathrm{CsV_3Sb_5}}}
\newcommand{\nstcdw}{$T_{\text{CDW}}$}
\newcommand{\avs}{\ensuremath{\mathrm{AV_3Sb_5}} }

\unnumbered



\begin{document}

\title[Article Title]{Soft Mode Origin of Charge Ordering in Superconducting Kagome \ch{CsV3Sb5}}

\author[1]{\fnm{Philippa Helen} \sur{McGuinness}}
\equalcont{These authors contributed equally to this work.}

\author[1]{\fnm{Fabian} \sur{Henssler}}
\equalcont{These authors contributed equally to this work.\vspace*{-0.2cm}}

\author[2,3]{\fnm{Manex} \sur{Alkorta}}

\author[1,4]{\fnm{Mark Joachim} \sur{Graf von Westarp}}

\author[5]{\fnm{Artem} \sur{Korshunov}}

\author[5]{\fnm{Alexei} \sur{Bosak}}

\author[6,7]{\fnm{Daisuke} \sur{Ishikawa}}

\author[6,7]{\fnm{Alfred Q. R.} \sur{Baron}}

\author[1,8]{\fnm{Michael} \sur{Merz}}

\author[1]{\fnm{Amir-Abbas} \sur{Haghighirad}}

\author[9, 10,11]{\fnm{Maia G.} \sur{Vergniory}}

\author[1]{\fnm{Sofia-Michaela} \sur{Souliou}}

\author[1]{\fnm{Rolf} \sur{Heid}}

\author[2,3,9]{\fnm{Ion} \sur{Errea}}

\author*[1]{\fnm{Matthieu} \sur{Le Tacon}}\email{matthieu.letacon@kit.edu}

\affil[1]{\orgdiv{Institute for Quantum Materials and Technologies}, \orgname{Karlsruhe Institute of Technology}, \city{Karlsruhe}, \postcode{76021}, \country{Germany}}

\affil[2]{Centro de Física de Materiales (CFM-MPC), CSIC-UPV/EHU, Donostia, 20018, Spain}

\affil[3]{Department of Applied Physics, University of the Basque Country (UPV/EHU), Donostia, 20018, Spain}

\affil[4]{\orgdiv{Max Planck Institute for Solid State Research, Heisenbergstraße 1, D-70569 Stuttgart, Germany}}

\affil[5]{\orgdiv{ESRF, The European Synchrotron, 71, avenue des Martyrs, CS 40220 F-38043 Grenoble Cedex 9}}

\affil[6]{Materials Dynamics Laboratory, RIKEN SPring-8 Center, Kouto 1-1-1, Sayo Hyogo 679–5148, Japan}

\affil[7]{Precision Spectroscopy Division, SPring-8/JASRI,
%Japan Synchrotron Radiation Research Institute, 
Kouto 1-1-1, Sayo, Hyogo 679–5198, Japan}

\affil[8]{Karlsruhe Nano Micro Facility (KNMFi), Karlsruhe Institute of Technology, Kaiserstr. 12, 76131 Karlsruhe, Germany}

\affil[9]{Donostia International Physics Center (DIPC), Donostia, 20018, Spain}

\affil[10]{Département de Physique et Institut Quantique, Université de Sherbrooke, Sherbrooke, J1K 2R1 Québec, Canada}

\affil[11]{Regroupement Québécois sur les Matériaux de Pointe (RQMP), Québec H3T 3J7, Canada}

\abstract{Charge-density-wave (CDW) order and superconductivity coexist in the kagome metals \ch{AV3Sb5}~(A=K, Cs, Rb), raising fundamental questions about the mechanisms driving their intertwined phases. Here we combine high-resolution inelastic X-ray scattering with first-principles calculations to uncover the origin of CDW formation in \ch{CsV3Sb5}. Guided by structure factor analysis, we identify a soft phonon mode along the reciprocal \textit{M}-\textit{L} direction, with the strongest effect at the \textit{L} point, where the elastic scattering intensity also grows most rapidly upon cooling. First-principles calculations incorporating lattice anharmonicity and electron-phonon coupling reproduce these observations and establish a soft-mode instability at the \textit{L} point as the driving mechanism of CDW formation. Despite the weakly first-order character of the transition, our results unambiguously demonstrate that the CDW in \ch{CsV3Sb5} originates from a softened phonon, clarifying its microscopic origin and highlighting the central role of lattice dynamics in kagome metals.}


\maketitle

The two-dimensional kagome lattice, built from corner sharing triangles and long recognized as a paradigmatic frustrated geometry~\cite{syozi1951statistics}, has emerged as a fertile ground for exotic quantum phenomena ~\cite{ghimire2020topology,kiesel2013}.
The recently discovered kagome metals \avs(A=K, Cs, Rb) have therefore sparked intense interest. Their electronic structure is remarkably rich, hosting Dirac cones~\cite{hao2022dirac}, flat bands~\cite{ortiz2019new}, multiple van Hove singularities~\cite{kiesel2013,kang2022twofold} and even a non-trivial $\mathbb{Z}_2$ topological invariant~\cite{ortiz2020cs}. On top of this unusual band topology, these compounds exhibit a charge-density-wave (CDW) transition, at $T_{\text{CDW}}\approx$~\SI{94}{K} for \nscvs, followed, at lower temperature, by superconductivity with $T_c=$~\SI{2.5}{K}~\cite{ortiz2020cs}. Although the coexistence of a CDW and superconductivity is not unique to kagome metals, in this family it acquires a particularly intriguing character, providing a compelling platform to explore how geometry, topology, and electronic correlations conspire to produce intertwined quantum orders.

Despite substantial effort, however, the nature and origins of the CDW in \cvs remain heavily disputed. Both 2$\times$2$\times$2~\cite{liang2021three,li2021observation,zhao2021cascade} and 2$\times$2$\times$4~\cite{ortiz2021fermi} superstructures have been reported as well as coexistence~\cite{xiao2023coexistence} and transitions between these orders as a function of temperature~\cite{stahl2022temperature,kautzsch2023structural} and even as a function of the sample cooling rate~\cite{xiao2023coexistence}. The CDW transition temperature and wavevector, as well as the superconductivity, also exhibit a strong sensitivity to tuning parameters such as pressure~\cite{stier2024pressure,zheng2022emergent,yu2021unusual,feng2023commensurate} and chemical substitution~\cite{oey2022fermi,liu2023doping,yang2022titanium,liu2022evolution,zhou2023effects,ding2022effect}. 

In addition, the mechanism behind the formation of the CDW remains unknown. A CDW can be driven by a Peierls-like instability determined by Fermi surface nesting~\cite{kohn1959image,rice1975new} or correlation, such as momentum-dependent electron-phonon coupling (EPC)~\cite{johannes2008fermi,zhu2015classification}. In \nscvs, assuming a nearest-neighbor tight-binding model, the CDW wavevector coincides with the nesting vector between reciprocal \textit{M}~points, which host nearby van Hove singularities (VHSs)~\cite{kang2022twofold}. Therefore, some studies have suggested that the nesting of the \textit{M} points is responsible for the formation of the CDW~\cite{tan2021charge,deng2023two,zhou2021origin,jiang2021unconventional,denner2021analysis,jin2022interplay}. Others, however, considering that the fermiology is not that simple, have proposed that EPC is likely responsible~\cite{kaboudvand2022fermi,liu2022observation,kaboudvand2022fermi,alkorta2025symmetrybrokengroundstatephonon,gutierrez2024phonon}. In fact, ARPES~\cite{luo2022electronic,azoury2023direct}, Raman scattering~\cite{liu2022observation, wu2022charge, He2024Raman} and infrared spectroscopy~\cite{uykur2022optical} studies have all found signatures of substantial EPC in these materials. 

Multiple harmonic calculations of the phonon dispersion have shown that the high symmetry phase of CsV$_3$Sb$_5$ is dynamically unstable with an instability along the \textit{M}-\textit{L} direction~\cite{tan2021charge,ratcliff2021coherent,wu2022charge}. In principle, inelastic scattering should offer an ideal method to investigate whether specific phonon branches are indeed unstable at these or other wave vectors. However, several studies have found no unambiguous evidence for sizable phonon anomalies associated with the CDW formation. In particular, neither inelastic X-ray scattering (IXS) studies~\cite{li2021observation, subires2023order} nor an inelastic neutron scattering study~\cite{xie2022electron} found evidence of phonon anomalies in \avs above \nstcdw. 
Recent theoretical work has shown that the stability of the high-symmetry phase can only be understood by accounting for ionic entropy and anharmonicity, suggesting that, despite the weakly first-order character of the CDW, a strong phonon renormalization occurs in the system~\cite{alkorta2025symmetrybrokengroundstatephonon} and that the substantial broadening of the soft phonon due to scattering might render it unobservable~\cite{gutierrez2024phonon}.

To resolve this apparent contradiction and uncover the mechanism underlying CDW formation in \ch{CsV3Sb5}, we performed a new series of experiments guided by structure factor calculations. These revealed a pronounced softening of a phonon branch along the entire \textit{M}-\textit{L} direction in reciprocal space, with the strongest effect at the \textit{L} point --- where the associated elastic intensity also grows most rapidly upon cooling~\cite{subires2023order}. First-principles calculations, incorporating ionic fluctuations and a non-perturbative treatment of anharmonicity, reproduce these observations along with the correct transition temperature~\cite{alkorta2025symmetrybrokengroundstatephonon}. Together, our results provide unambiguous evidence that the CDW in CsV$_3$Sb$_5$ is driven by a phonon instability centered at the \textit{L} point and emphasize the pivotal role of quantum anharmonic effects in shaping the remarkable phase diagram of these kagome materials.

%\section{Results}\label{sec2}

\begin{figure}[h]
    \centering
    \includegraphics[width=0.6\linewidth]{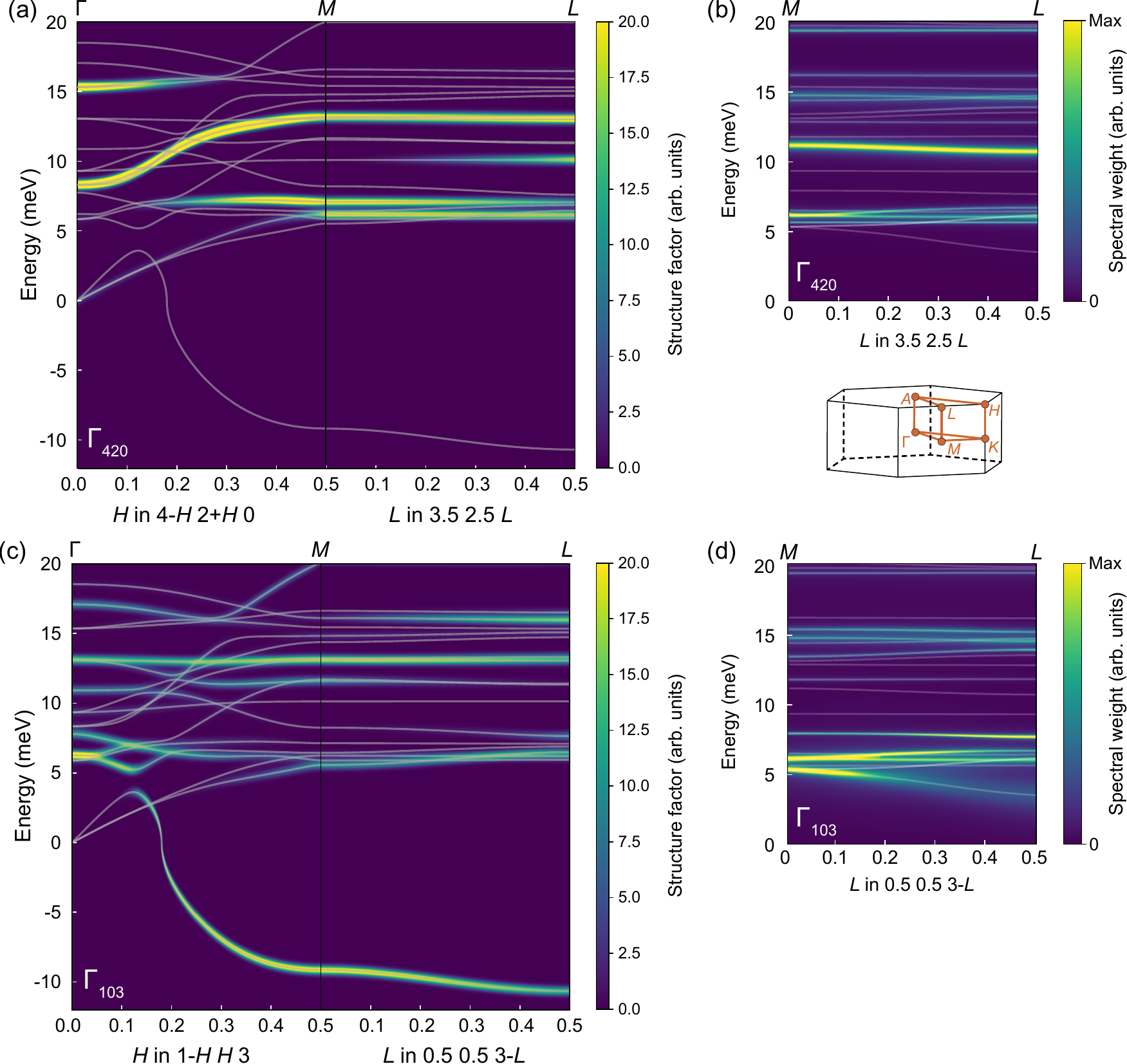}
    \caption{ Calculated phonon intensities in $\Gamma_{420}$ and $\Gamma_{103}$ for both harmonic and anharmonic calculations. (a,c) The calculated dynamic structure factor from harmonic DFPT calculations along the line $\Gamma$-\textit{M}-\textit{L} in a) $\Gamma_{420}$ c) $\Gamma_{103}$. The harmonic phonon frequencies are marked with a thin white line. (b,d) Spectral weight along the line \textit{M}-\textit{L} as calculated within the SSCHA formalism at 100 K (see Supplementary Information for details), which includes both the structure factor as well as a realistic broadening of the phonon peaks due to anharmonic and electron-phonon interactions, for b) $\Gamma_{420}$ d) $\Gamma_{103}$. The position of the expected peak of the spectral function is marked with a thin white line as a reference.}
    \label{fig:StructureFactor}
\end{figure}

We begin by examining the harmonic phonon spectrum of the high-temperature $P6/mmm$ phase of CsV$_3$Sb$_5$, calculated using density functional perturbation theory (DFPT) and based on the experimental room-temperature lattice parameters (see Ref.~\citenum{frachet2024colossal} for details). Consistent with earlier studies~\cite{tan2021charge,ratcliff2021coherent,subedi2022Hexagonal}, the phonon dispersion exhibits instabilities near the high-symmetry \textit{M} and \textit{L} points, reflected by imaginary frequencies plotted as negative values in Fig.~\ref{fig:StructureFactor}.
One branch displays a pronounced instability along the entire \textit{M--L} direction. This anomalous behavior forms the central focus of our experimental investigation. The inelastic scattering intensity associated with a given phonon mode varies strongly across different Brillouin zones (BZ). To optimize the experimental geometry and ensure sufficient signal for the mode of interest, we calculated the dynamical structure factor in all experimentally-accessible zones to identify the most favorable scattering conditions.
The results of these calculations are shown in Fig.~\ref{fig:StructureFactor}a,c for two representative BZ centered at $\Gamma_{420}$ and $\Gamma_{103}$, respectively, with the full phonon dispersion overlaid in white. 

The point $\mathbf{Q} = (3.5,\ 2.5,\ 0.5)$ corresponds to an \textit{L}~point within the $\Gamma_{420}$ zone, where previous IXS measurements have been performed~\cite{subires2023order}. However, the structure factor of the unstable branch at this location is vanishingly small, suggesting that the anomalous phonon mode would be essentially unobservable there.  By contrast, in the $\Gamma_{103}$ zone, the structure factor is substantial at both the \textit{M} point $\mathbf{Q} = (0.5,\ 0.5,\ 3)$ and the \textit{L} point $\mathbf{Q} = (0.5,\ 0.5,\ 2.5)$, indicating favorable conditions for detecting this mode. Notably, such favorable scattering conditions are rare: fewer than 10\% of BZs exhibit an \textit{L}~point structure factor for this branch that reaches even half the intensity calculated at $\mathbf{Q} = (0.5,\ 0.5,\ 2.5)$. These conclusions are robust against the inclusion of anharmonic effects, which stabilize the otherwise unstable phonon branches above $T_{\text{CDW}}$~\cite{gutierrez2024phonon,alkorta2025symmetrybrokengroundstatephonon}, and the impact of the electron-phonon scattering on the phonon linewidth. This is illustrated in Fig.~\ref{fig:StructureFactor}b,d, where we show the phonon spectral weight along the \textit{M}–\textit{L} direction in the two Brillouin zones considered, explicitly accounting for both anharmonicity and electron-phonon interactions. 

\begin{figure}[h]
    \centering
    \includegraphics[width=1\linewidth]{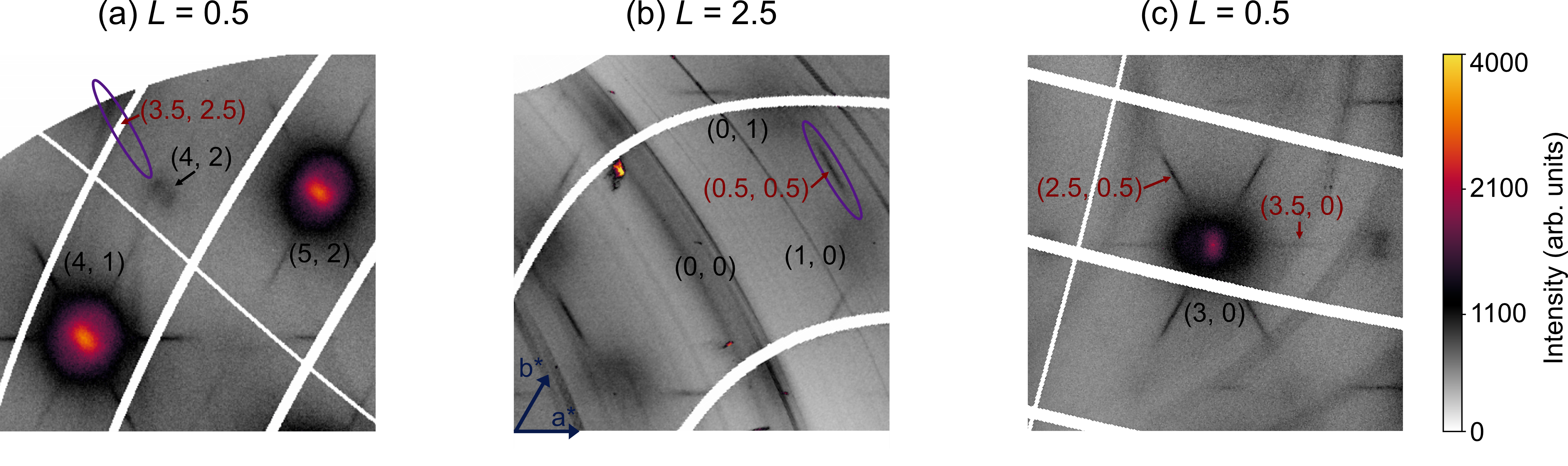}
    \caption{ Thermal diffuse scattering just above $T_{\text{CDW}}$. (a-c) The thermal diffuse scattering at $\SI{95}{K}$ in (a) $\Gamma_{420}$ (b) $\Gamma_{103}$, where all of the facets of the hexagon are crystallographically equivalent, (c) $\Gamma_{300}$. Some \textit{A} and \textit{L} points are indicated in black and red text, respectively.}
    \label{fig:TDS}
\end{figure}

The importance of BZ selection is further confirmed by thermal diffuse scattering (TDS) measurements. Fig.~\ref{fig:TDS} shows parts of the reconstructed (\textit{H}~\textit{K}~0.5) and (\textit{H}~\textit{K}~2.5) planes at \SI{95}{K}, i.e. just above $T_{\mathrm{CDW}}$. As previously reported~\cite{subires2023order}, diffuse streaks along the \textit{A--L} direction (such as indicated by the purple oval in Fig.~\ref{fig:TDS}b) emerge below \SI{150}{K}, with intensity peaking at \textit{L} and sharpening into super-lattice peaks at the CDW transition.

The intensity of the diffuse precursors to the CDW satellites vanishes in certain BZs. For instance, it is absent in the \textit{A--L} path from $\mathbf{Q} = (4,\ 2,\ 0.5)$ to $\mathbf{Q} = (3.5,\ 2.5,\ 0.5)$ (purple oval in Fig.~\ref{fig:TDS}a) — in agreement with the negligible structure factor for the unstable branch at the $\Gamma_{420}$ zone (Fig.~\ref{fig:StructureFactor}a) and the absence of phonon anomalies at this \textit{L} point~\cite{subires2023order}. Similarly, the diffuse streak is only weakly visible around the \textit{L} point at $\mathbf{Q} = (3.5,\ 0,\ 0.5)$ (Fig.~\ref{fig:TDS}c), where no phonon anomalies are observed at the corresponding \textit{M} point~\cite{li2021observation}. By contrast, substantial diffuse intensity is observed around the \textit{L} point $\mathbf{Q} = (2.5,\ 0.5,\ 0.5)$; however no phonon softening was detected at this point in previous IXS experiments conducted in a lower-resolution (\SI{3}{meV}) configuration~\cite{subires2023order}. 

Notably, the enhanced precursor signal observed around the \textit{L} point at $\mathbf{Q} = (0.5,\ 0.5,\ 2.5)$ (Fig.~\ref{fig:TDS}b), together with the favorable structure factor for the unstable branch in the corresponding $\Gamma_{103}$ zone (Fig.~\ref{fig:StructureFactor}b), suggests increased inelastic spectral weight relative to other BZs, motivating our targeted inelastic measurements for definitive confirmation.

\begin{figure}[h]
    \centering
    \includegraphics[width=1\linewidth]{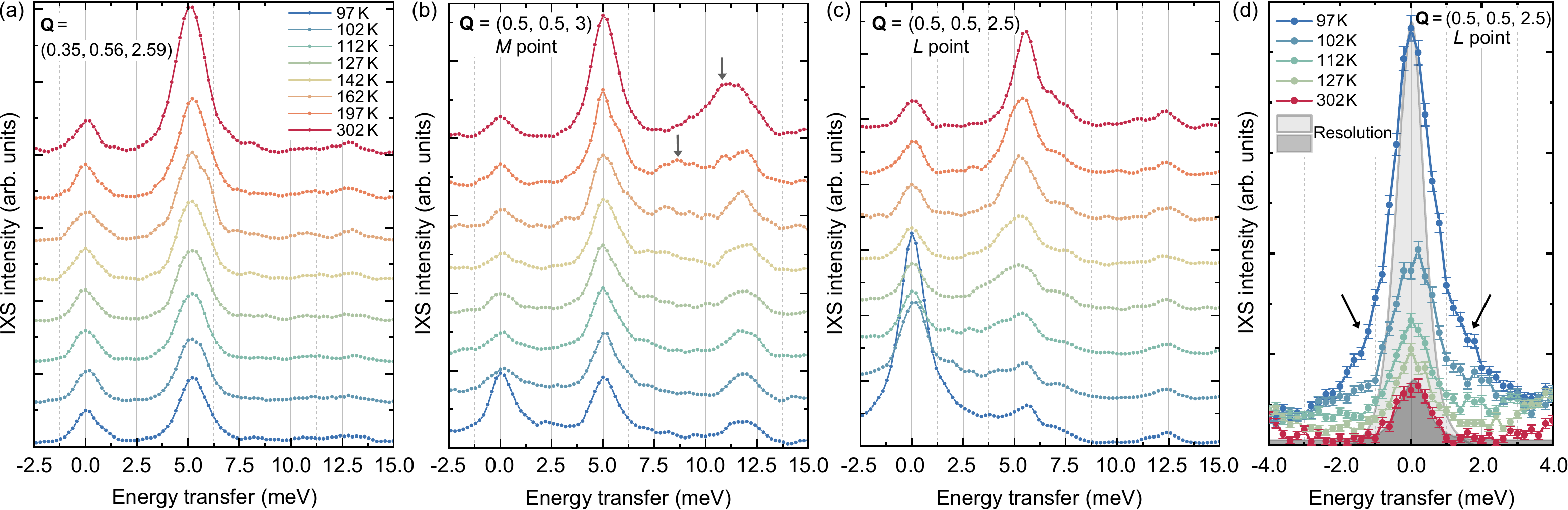}
    \caption{Raw IXS spectra. (a-c) Non-normalized IXS spectra, vertically offset for clarity, measured at temperatures between $\SI{97}{K}$ and $\SI{300}{K}$ at (a) a low-symmetry point $\mathbf{Q} = (0.35,\ 0.56,\ 2.59)$  (b) the \textit{M}~point $\mathbf{Q} = (0.5,\ 0.5,\ 3)$  (c) the \textit{L}~point $\mathbf{Q} = (0.5,\ 0.5,\ 2.5)$. The y-axis scale is not common across the subfigures. (d) A magnified, non-offset view of the low-energy range of (c) with overlays of the measured instrumental resolution scaled separately to the $\SI{302}{K}$ and $\SI{97}{K}$ data. Arrows indicate areas of enhanced low-energy spectral weight at 97~K which are not consistent with only a resolution-limited elastic peak.}
    \label{fig:RawSpectra}
\end{figure}

Fig.~\ref{fig:RawSpectra}a--c shows raw, high-resolution (1.1 meV)~\cite{ishikawa2010temperature,Baroninprep} IXS spectra recorded between \SI{97}{K} and \SI{300}{K} at a low-symmetry point $\mathbf{Q} = (0.35,\ 0.56,\ 2.59)$ and at the \textit{M} point $\mathbf{Q} = (0.5,\ 0.5,\ 3)$ and the \textit{L} point $\mathbf{Q} = (0.5,\ 0.5,\ 2.5)$ in the $\Gamma_{103}$ zone. The low-symmetry point shows no anomalies beyond the expected reduction in phonon intensity with temperature due to the Bose factor. In contrast, the \textit{L} point exhibits a clear increase in elastic peak intensity upon cooling, with a similar but weaker effect at the \textit{M} 
point.

At the \textit{M} point, additional changes appear in the $\SI{7}{meV} - \SI{12}{meV}$ range: a broad peak around $\SI{11}{meV}$ at $\SI{300}{K}$ (arrow) appears to split at lower temperatures, indicating a phonon mode evolution. Upon approaching $T_{\text{CDW}}$, the low energy spectral weight is enhanced at the \textit{M} point and -- even more so -- at the \textit{L} point. Thanks to the high resolution, we can confirm that this enhancement exceeds a mere increase in the resolution-limited elastic scattering intensity (Fig.~\ref{fig:RawSpectra}d), consistent with the presence of a softening phonon mode. In lower resolution measurements, by contrast, the inelastic contribution to this spectral weight enhancement is less clearly discernible (see Supplementary Information).

\begin{figure}[h]
    \centering
    \includegraphics[width=\linewidth]{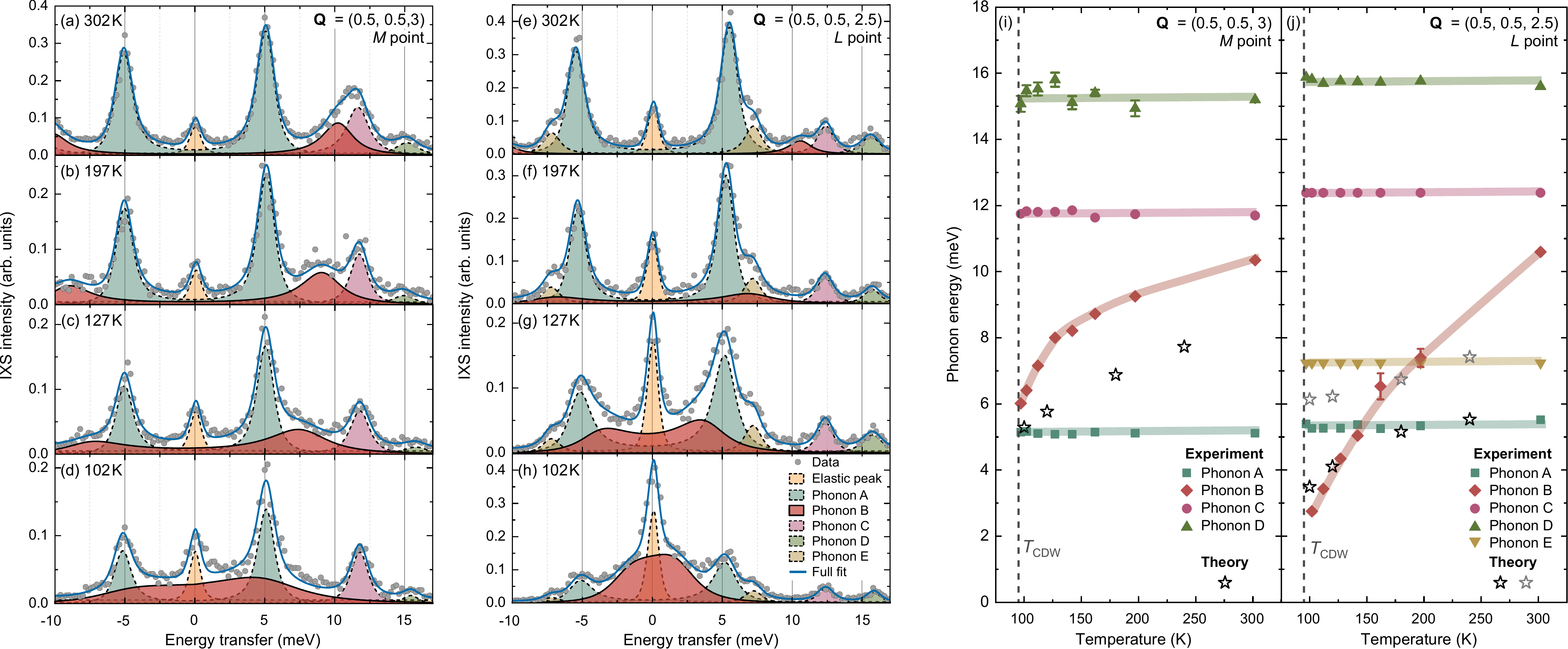}
    \caption{Fitted IXS spectra and phonon energies at the \textit{M} and \textit{L} points as a function of temperature. The y-axis scale is not common among the panels. (a-h) Fitted IXS spectra at the (a-d) \textit{M} and (e-h) \textit{L} points at (a,e) $\SI{302}{K}$, (b,f) $\SI{197}{K}$, (c,g) $\SI{127}{K}$ and (d,h) $\SI{102}{K}$ showing individual phonon modes, the elastic line and the overall sum of all of these contributions (full fit). (i,j) The fitted phonon energies as a function of temperature for (i) the \textit{M}~point (j) the \textit{L}~point alongside calculated anharmonic values for the modes which show softening with temperature (stars). The lines represent guides to the eye.}
    \label{fig:FittedSpectra}
\end{figure}

In order to provide more insight as to the precise nature of these spectral changes, we fit the data shown in Fig.~\ref{fig:RawSpectra} using the measured resolution function for the elastic peak and a set of damped harmonic oscillators (DHOs) for the phonons, convoluted with the resolution.
The fitting approach and the functions used are described in more detail in the Methods section. The individual fitted phonons and elastic line are shown for a series of temperatures at the \textit{M} (Figs.~\ref{fig:FittedSpectra}a-d) and \textit{L}~points (Figs.~\ref{fig:FittedSpectra}e-h). At the \textit{M}~point (resp. \textit{L}~point), there are four (resp. five) fitted phonons.

As shown in Figs.~\ref{fig:FittedSpectra}i-j, most fitted phonon branches are essentially temperature independent, with no discernible change in energy or linewidth beyond the experimental uncertainty. A notable exception occurs at the \textit{M} point (Fig.~\ref{fig:FittedSpectra}i), where a branch at $\SI{10}{meV}$ at $\SI{302}{K}$ softens dramatically to $\SI{6}{meV}$ at $\SI{97}{K}$, just above \nstcdw. The effect is even stronger at the \textit{L} point (Fig.~\ref{fig:FittedSpectra}j), where  the same branch softens below $\SI{3}{meV}$ at $\SI{102}{K}$, representing an exceptionally large renormalization. At $\SI{97}{K}$, it is not possible to reliably extract the energy of this mode, which becomes overdamped.

Such strong temperature dependence is only reproduced when anharmonic effects are explicitly included in lattice dynamics calculations.
In solids, anharmonicity gives rise to thermal expansion and governs the phonon-phonon interactions responsible for the temperature-dependent frequency shifts and linewidth changes, in addition to the temperature-independent effects of EPC. These are not included in DFPT, which treats phonons within a harmonic framework at $T=\SI{0}{K}$. They can, however, be incorporated non-perturbatively using the stochastic self-consistent harmonic approximation (SSCHA)~\cite{Errea2014Anharmonic,monacelliStochasticSelfconsistentHarmonic2021}, albeit at a significantly higher computational cost. Using this approach, we find that most phonons remain temperature independent (see Supplementary Information), with the notable exceptions of a \textit{$M_1^+$} and two \textit{$L_2^-$} modes, shown in Figs.~\ref{fig:FittedSpectra}i-j. The \textit{$M_1^+$} and one of the two \textit{$L_2^-$} modes belong to the branch which is unstable in the harmonic calculation but is stabilized by anharmonicity (see also Fig.~\ref{fig:StructureFactor}-c and d) and which displays strong softening upon cooling~\cite{gutierrez2024phonon}, consistent with the experiment. The theory also predicts an avoided crossing (see Fig. ~\ref{fig:FittedSpectra}j and Supplementary Information) and transfer of spectral weight between two \textit{$L_2^-$} modes at about \SI{5}{meV}. The avoided crossing is absent by symmetry at \textit{M}. Due to the overlapping phonons in this energy region and the experimental resolution, however, fitting cannot capture the anti-crossing, most pronounced in the vicinity of \textit{L}.

\begin{figure}[h]
    \centering
    \includegraphics[width=\linewidth]{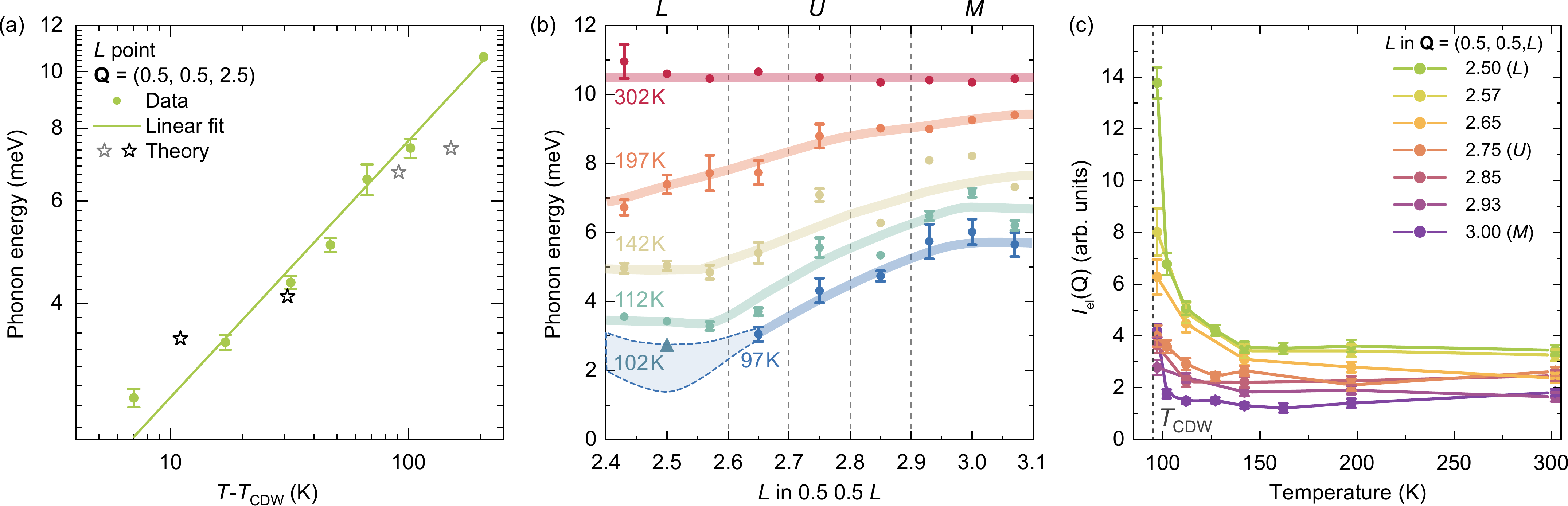}
    \caption{Order of the transition, scale of the softening effect in reciprocal space and growth of the elastic peak. (a) A log-log plot of the energy of Phonon~B at the \textit{L} point as a function of the reduced temperature  $T-$\nstcdw, along with calculated values for the softening mode before and after the avoided crossing (stars). (b) The energy of Phonon~B along the high symmetry line  $\mathbf{Q} = (0.5,\ 0.5,\ L)$ as a function of temperature. The lines are guides to the eye. Some points are omitted at $\SI{97}{K}$ due to the phonon becoming overdamped, with data from 102~K being shown instead for the \textit{L}~point (triangle). The shaded region indicates approximate boundaries for the energy at 97~K for these points. (c) The intensity of the fitted elastic peak as a function of temperature along the line  $\mathbf{Q}~=~(0.5,\ 0.5,\ L)$.}
    \label{fig:MLline}
\end{figure}

In order to examine the temperature dependence of the energy of the soft branch in more detail, Fig.~\ref{fig:MLline}a shows the log-log energy dependence of this branch as a function of the relative temperature $T-$\tcdw for both the experimental data and the anharmonic calculations. There is a good agreement between the data and a linear fit where the fitted gradient is 0.44~$\pm$~0.02. Due to the proximity of this value to 0.5, it is impossible for us to completely rule out that the CDW transition shows mean field, second-order-like behavior from this measurement alone, but we note that our measurements of \tcdw from the zero-energy intensity, shown in the Supplementary Information, do show a small hysteresis ($\sim \SI{1}{K}$), in line with the weakly first-order behavior observed in thermodynamic measurements~\cite{kountz2024thermal,frachet2024colossal,ortiz2020cs}, also supported by free energy calculations including lattice anharmonicity~\cite{alkorta2025symmetrybrokengroundstatephonon}. 

Having established the existence of a phonon anomaly at the \textit{M} and \textit{L} points, we now discuss the dispersion of this phonon along the \textit{M}-\textit{L} line  $\mathbf{Q} = (0.5,\ 0.5,\ L)$, as shown in Fig.~\ref{fig:MLline}b. At $\SI{302}{K}$, the phonon is essentially non-dispersive, with a momentum-independent energy of around $\SI{10.3}{meV}$. Even at $\SI{197}{K}$, however, there is a clear dispersive effect: not only has the phonon softened everywhere along this line, but the degree of softening increases essentially monotonically from the \textit{M} to the \textit{L} point. As the temperature decreases towards $T_{\text{CDW}}$, the phonon continues to soften along the whole line, but the dispersive effect remains and strengthens. This effect confirms that the \textit{L} point is the leading lattice instability in the system but also reveals that the phonon anomaly is unusually broad in reciprocal space. Additionally, in order to gauge the size of the anomaly in-plane, we also performed measurements from the \textit{M} point $\mathbf{Q} = (0.5,\ 0.5,\ 3)$ towards the $\Gamma$ point $\mathbf{Q} = (1,\ 0,\ 3)$, which reveal that the softening effect gradually decreases away from \textit{M} and fully disappears by $\mathbf{Q} = (0.75,\ 0.25,\ 3)$, halfway between $\Gamma$ and \textit{M}, as detailed in the Supplementary Information. 

As is clear even from the raw spectra in Fig.~\ref{fig:RawSpectra}, with decreasing temperature, as well as the growing contribution to the low-energy spectral weight due to the softening phonon, the elastic scattering is also enhanced. Fig.~\ref{fig:MLline}c shows the fitted elastic scattering intensity, $I_{\text{el}}(Q)$, as a function of temperature for $Q$-points along the line between \textit{M} and \textit{L} (see TDS maps in Supplementary Information). The intensity increases for all points with decreasing temperature, but most strongly in proximity to \textit{L}. This enhancement signals the development of quasi-static CDW correlations. As the transition is approached, these fluctuations slow down and grow in spatial extent, giving rise to resolution-limited scattering that evolves into superlattice Bragg peaks below $T_{\mathrm{CDW}}$.

Taken together, our experimental and theoretical results demonstrate that the CDW is driven by the softening of an \textit{L$_{2}^{-}$} phonon, identifying the \textit{L} point as the leading instability and favoring a three-dimensional 2$\times$2$\times$2 ordering pattern. They further rule out that the CDW transition in \cvs is driven by a nesting of the van Hove singularities at the \textit{M} point~\cite{tan2021charge,deng2023two,zhou2021origin,jiang2021unconventional,denner2021analysis,jin2022interplay,kiesel2013,gutierrez2024phonon}. 
The phonon softening effect is broad in reciprocal space: the phonon softens by at least $\SI{4}{meV}$ everywhere along the \textit{M-L} line and the softening continues to halfway between $\Gamma$ and \textit{M} in-plane, encompassing half the BZ and closely matching the region where calculations predict strong EPC in the unstable branch (Supplementary Information). This scale suggests a correlation driven effect rather than a Kohn-like anomaly~\cite{zhu2015classification} and is more similar to established anisotropic EPC-driven CDW materials such as 2\ch{H-NbSe_2},  \ch{TbTe_3}, and 1\ch{T-VSe_2}~\cite{weber2011extended,maschek2015wave,diego_van_2021} than systems which host a Kohn anomaly~\cite{hoesch2009giant,renker1973observation, Bosak_PRR2021}.

This conclusion is also in agreement with hydrostatic pressure studies, which found a high sensitivity of the CDW wavevector and ordering temperature to pressure, even at pressures below 1~GPa, despite calculations based on the high pressure structure suggesting there is little change to the van Hove singularities or the nesting~\cite{stier2024pressure}, and a time-resolved ARPES study which found evidence for EPC being the main mechanism for the CDW formation~\cite{azoury2023direct}.

We note that, in both the calculation and experimental data, the modes with the highest sensitivity are those belonging to the \textit{M}$_{1}^+$/\textit{L}$_{2}^-$ irreducible representations. Modes without this symmetry show little temperature dependence and, as shown in the Supplementary Information, only extremely weak electron-phonon coupling. In addition, the \textit{M}$_{1}^+$/\textit{L}$_{2}^-$ modes are most sensitive to the details of the calculation and parameters such as the electron temperature in the purely harmonic case~\cite{gutierrez2024phonon,tan2021charge,deng2023two}.

As discussed above, the unstable phonon near the \textit{L} point becomes overdamped at low temperature, with a fitted linewidth exceeding its damped energy. In this regime, the DHO model no longer provides a reliable description. Although a Lorentzian line shape is sometimes employed in such cases~\cite{yamada1969study}, the convolution with the instrumental resolution prevented us from separating these contributions here, and we therefore omitted the corresponding points from Figs.~\ref{fig:FittedSpectra} and \ref{fig:MLline}. 

Reliable fitting close to $T_{\text{CDW}}$ is further hindered by the rapid growth of the elastic peak below 110 K, which dominates the spectra near the \textit{L} point and obscures the soft-phonon linewidth. By contrast, at the \textit{U} and \textit{M} points, where the elastic line is weaker, the fits are more robust and reveal that the inelastic spectra cannot be described without a substantial and continuous broadening of the soft branch (see Supplementary Information). This pronounced linewidth increase is also reproduced by first-principles calculations and originates from the combined effects of strong EPC and lattice anharmonicity. These results highlight that both play a central role in shaping the low-energy physics of kagome metals, further supporting as well the phonon origin of superconductivity in \cvs\cite{alkorta2025symmetrybrokengroundstatephonon}.

Beyond identifying the nature and microscopic origin of the CDW in \nscvs, our findings establish the central role of lattice dynamics in governing the electronic instabilities of kagome metals. The strong coupling between electronic and lattice degrees of freedom demonstrated here provides a natural framework to understand how superconductivity, topology, and charge order emerge and interact in these materials. More broadly our work underscores the importance of phonon-driven instabilities in geometrically frustrated metals and paves the way for a unified understanding of correlated kagome systems, where intertwined orders emerge from the delicate balance between electronic correlations, topology and lattice anharmonicity.

\section*{Methods and Materials}

\subsection{Crystal growth}
High-purity elements Cs (Alfa Aesar, 99.98\%), V (Cerac/Pure, 99.9\%, further purified to remove oxygen), and Sb (Gmaterials, 99.999\%) were mixed in a molar ratio of 2:1:6 in an alumina crucible, sealed inside an iron container under argon atmosphere. The container was placed in a tube furnace (argon-sealed), heated to \SI{1050}{^\circ C} and held for \SI{15}{h}. The melt was then cooled to \SI{650}{^\circ C} at \SI{1.65}{^\circ C/h}, at which point the furnace was tilted to decant excess flux, followed by cooling to room temperature. The crystals were washed in demineralized water to remove residual flux.
The resulting single crystals were characterized prior to the IXS/TDS experiments. Their chemical composition was examined by energy-dispersive X-ray spectroscopy (EDS) using a COXEM EM-30plus electron microscope equipped with an Oxford Silicon-Drift Detector and the AZtecLiveLite software. The EDS analysis yielded an elemental composition of Cs:V:Sb = 1:3:5.1. For samples from each batch, \tcdw and $T_\text{c}$ were verified via magnetic susceptibility measurements. Structural refinement and thermodynamic characterization of samples grown via the same protocol can be found in Refs.~\citenum{stier2024pressure} and~\citenum{frachet2024colossal}.

\subsection{TDS experiment}
The thermal diffuse X-ray scattering experiments were performed at beamline ID28 of the European Synchrotron Radiation Facility (ESRF) \cite{Krisch2007,Girard2019}. The incident X-ray beam energy was set to \SI{17.794}{\kilo\eV} and the beam was focused to a spot of approximately \qtyproduct[product-units=power]{40 x 40}{\micro\meter}. The data were acquired with a Pilatus3 X 1M detector in shutterless mode while continuously rotating the sample and recording images while integrating over an angular range of \ang{0.25} and \SI{0.5}{\second}. The CrysAlis Pro software package~\cite{CrysalysPro} was used for determining the unit cell and sample orientation. The reciprocal space reconstructions were created with a software developed at the beamline ID28. Low-temperature conditions were achieved using an Oxford Cryostream 700 Plus cooling system. 

\subsection{IXS experiments}
The IXS experiments were performed at beamline ID28~\cite{Krisch2007,Girard2019} of the ESRF in July 2023 and at beamline BL43LXU~\cite{Baron2010,Baron2015} of the RIKEN SPring-8 Center (Japan) in November 2024.
At ID28, the incident X-ray beam energy was set to \SI{17.794}{\kilo\eV} using the \((9\ 9\ 9)\) silicon reflection with an energy resolution of \SI{3}{\milli\eV}. The beam was focused to a spot of \qtyproduct[product-units=power]{25 x 25}{\micro\meter} and the momentum resolution was set to \(\approx\SI{0.25}{\per\nano\meter}\) in the scattering plane and \(\approx\SI{0.75}{\per\nano\meter}\) perpendicular to it. Low-temperature conditions were achieved using an Oxford Cryostream 700 Plus cooling system.

At BL43LXU, the IXS spectrometer was operated at an X-ray energy of \SI{23.724}{keV} using the silicon \((12\ 12\ 12)\) backscattering reflection with a nominal \SI{1.1}{meV}~\cite{ishikawa2010temperature,Baroninprep} energy resolution. 
A two-dimensional array of $7\times4=28$ analyzers was used to collect data from 28 different momentum transfers in each scan.  The energy resolution of each analyzer was determined as discussed in Ref.~\citenum{ishikawa2021practical} and was 1.1-\SI{1.2}{meV} for most of the analyzers for which data is presented.  The momentum resolution was set by $40\times$\SI{40}{mm\squared} slits at \si{9}{m} from the sample, corresponding to a momentum resolution of $\Delta\mathbf{Q} = (\Delta H,\ \Delta K,\ \Delta L) =(0.03,0.05,0.06)$ reciprocal lattice units, full width.  For measurements at low temperatures, the samples were mounted inside a closed-cycle cryostat.

\subsection{Fitting}

As described in the main text, the phonons were fitted using a damped harmonic oscillator function, $S_j(Q,E)$ which is a function of the wave vector $Q$ and the energy $E$, weighted by the Bose factor. For the $j$th phonon,

\begin{equation}
S_j(Q,E) = \frac{A_j }{\pi}\frac{1}{1 - e^{-\frac{E}{k_{\text{B}} T}}}  \frac{4  \Gamma_j E}{ \left( E^2 - (E_j^2 + \Gamma_j ^2) \right)^2 + 4 (\Gamma_j E)^2 },
\end{equation}

where $T$ is the temperature, $A_j$ is the phonon structure factor, $E_j$ the damped phonon energy of the $j$th phonon and $\Gamma_j$ is the damping rate (the HWHM of the phonon). The undamped phonon energy is then $E_{0j}=\sqrt{E_j^2 + \Gamma_j ^2}$, which is described simply as the phonon energy in the main text. 

Due to the challenges involved in fitting the softening phonon, Phonon~B, as it crosses over Phonon~A, the $A_j$ values for Phonons~A and B were fixed to their $\SI{300}{K}$ values. This assumption is supported by the calculations, which see a negligibly small change in this factor as a function of temperature at the \textit{M} point for the softening phonon and at the \textit{L} point when considering the sum of the structure factor for the branches involved in the avoided crossing. 

In addition, due to the proximity of Phonons B and C at $\SI{302}{K}$, to enable these phonons to be distinguished and not fitted as a single broad phonon, the width of Phonon~C was fixed to the fitted $\SI{197}{K}$ value during the fitting of the $\SI{302}{K}$ data -  as is clear from Fig.~\ref{fig:FittedSpectra}, this phonon shows no significant change in energy or width as a function of temperature, validating this procedure. 

The resolution for each analyzer was determined at the start of the experiment by measuring tempax glass, a mostly elastic scatterer, with the residual inelastic response of the glass removed as discussed in Ref.~\cite{ishikawa2021practical}, generating a sarf (smooth approximation to the resolution function). The sarf, in this case a combination of four Lorentzians and three Gaussians to accurately capture both the width and the tails of the resolution function, was used in the fitting procedure.

\subsection{Computational approach}

The temperature dependent quantum vibrational modes were computed within the stochastic self-consistent harmonic approximation (SSCHA) and its dynamical extension implemented in the SSCHA package~\cite{Errea2014Anharmonic,monacelliStochasticSelfconsistentHarmonic2021,BiancoSecondOrderStructural2017}. Considering the diagonal terms of the phonon-phonon interaction (Bubble approximation), along with linear response electron-phonon interaction, we approximated the phonon spectral function to a sum of individual Lorentzian distributions with finite linewidth. To compare with experiments, dynamic structure calculations were included. 
The force and energy calculations for the SSCHA calculations were performed with Gaussian approximation potential (GAP)~\cite{bartokGaussianApproximationPotentials2010} trained on density functional theory (DFT) data with the optB88-vdW~\cite{klimesChemicalAccuracyVan2009} functional for the exchange and correlation functional.
The electron-phonon linewidth was obtained from density functional perturbation theory (DFPT)~\cite{Baroni2001Phonons} as implemented in the Quantum Espresso package~\cite{Giannozzi_2009,Giannozzi_2017} making use of ultra-soft pseudopotentials~\cite{DALCORSO2014337}.
Each of the steps and assumptions made in the calculation are explained in detail in the Supplementary Information. 

For Fig \ref{fig:StructureFactor}(a,c) the harmonic phonons were computed in a separate framework. IXS structure factors were calculated on the basis of ab initio phonon frequencies and eigenvectors using the density-functional
perturbation theory as implemented in the mixed-basis pseudopotential method \cite{Heid1999}. Dynamical matrices were first calculated on a $6 \times 6 \times 2$ hexagonal mesh and then determined for arbitrary points in the Brillouin zone using standard Fourier-interpolation techniques, from which the phonon dispersion and structure factors were derived.
For this calculation, the computational parameters were the same as described in a previous publication (see supplemental material of Ref.~\citenum{frachet2024colossal}).

\backmatter
\bmhead{Supplementary information}
The online version contains supplementary material available at XXXXXX.

\bmhead{Acknowledgements}
We thank S. Blanco-Canosa for fruitful discussions. We acknowledge provision of beamtime at the RIKEN Quantum NanoDynamics Beamline, BL43LXU, under the proposal with number 20240083. We acknowledge the European Synchrotron Radiation Facility (ESRF) for provision of synchrotron  radiation facilities under proposal number HC-5341. We acknowledge the funding by the Deutsche Forschungsgemeinschaft (DFG; German Research Foundation) Project-ID 422213477-TRR 288 (Project B03 and B10).
R.H. acknowledges support by the state of Baden-W\"{u}rttemberg through bwHPC.  The work was also supported by the
PID2022-142861NA-I00 and PID2022-142008NB-I00 projects funded by MICIU/AEI/10.13039/501100011033 and FEDER, UE;
the Department of Education, Universities and Research of the Eusko Jaurlaritza and
the University of the Basque Country UPV/EHU (Grant
No. IT1527-22);
Canada Excellence Research Chairs
Program for Topological Quantum Matter;
NSERC Quantum Alliance France-Canada; and Diputación Foral de Gipuzkoa Programa Mujeres y Ciencia.
M.A. acknowledges a PhD scholarship from the Basque Government.

\bibliography{Kagomebib}
\end{document}